\documentclass[]{spie}  

\usepackage{amsmath,amsfonts,amssymb}
\usepackage{graphicx}
\usepackage{booktabs}
\usepackage[colorlinks=true, allcolors=blue]{hyperref}

\title{Comparing realtime optical gain measurement and methods on MagAO-X}

\author[a]{Eden McEwen}
\author[b]{Jared R. Males}
\author[a,b,c,d]{Olivier Guyon}
\author[b,f]{Sebastiaan Y. Haffert}
\author[c,e]{Vincent Deo}
\author[g]{Joseph D. Long}
\author[h]{Logan A. Pearce}
\author[b]{Laird M. Close}
\author[b]{Warren B. Foster}
\author[b]{Kyle Van Gorkom}
\author[i]{Alexander D. Hedglen}
\author[b]{Parker Johnson}
\author[b]{Maggie Y. Kautz}
\author[a]{Jay K. Kueny}
\author[b]{Jialin Li}
\author[a]{Joshua Liberman}
\author[b,c]{Miles Lucas}
\author[a]{Jennifer Lumbres}
\author[j]{Avalon L. McLeod}
\author[f]{Elena Tonucci}
\author[a]{Katie Twitchell}
\author[k]{Lauren Schatz}
\author[l]{Alycia J. Weinberger}

\affil[a]{Wyant College of Optical Sciences, The University of Arizona}
\affil[b]{Steward Observatory, The University of Arizona}
\affil[c]{Subaru Telescope, National Observatory of Japan}
\affil[d]{Astrobiology Center, National Institutes of Natural Sciences}
\affil[e]{Optical Sharpeners SAS, France}
\affil[f]{Leiden Observatory, Leiden University, The Netherlands}
\affil[g]{Center for Computational Astrophysics, Flatiron Institute}
\affil[h]{University of Michigan, Ann Arbor, MI}
\affil[i]{Northrop Grumman Corporation, Rolling Meadows, Illinois}
\affil[j]{Draper Laboratory, Cambridge, Massachusetts}
\affil[k]{Starfire Optical Range, Kirtland Air Force Base, Albuquerque, New Mexico}
\affil[l]{Earth \& Planets Laboratory, Carnegie Science, Washington DC}

\authorinfo{Further author information: (Send correspondence to E.A.M.)\\E.A.M.: E-mail: edenmcewen@arizona.edu}

\begin{document} 
\maketitle
\begin{abstract}
A lingering technical challenge for pyramid wavefront sensors (PyWFS) is their change in response between calibration and correction residuals, a quantity known as optical gain (OG). Given the prevalent use of PyWFSs in current and planned high contrast adaptive optics (AO), understanding and reliably measuring OG for realtime control unlocks advanced correction and post processing techniques. The OG quantity as an unknown inhibits a system’s ability to stably correct non common path errors, reconstructing wavefronts, and PSF reconstruction. This work compares kinds of optical gain measurement techniques on MagAO-X, a visible light extreme AO instrument on the 6.5m Magellan Clay telescope. We present a set of on-sky measurements of OG across three techniques: 1) An on-sky calibration that acquires OG per spatial mode, 2) realtime measurements of the instantaneous Strehl Ratio (SR) on the pyramid tip, and 3) realtime measurement of known, high-frequency probe signal on the WFS itself. We compare these on-sky results with performance diagnostics to asses how faithfully OG is returned. We conclude with future steps for active control of OG on MagAO-X.

\end{abstract}


\section{Introduction}
\label{sec:intro}  

Pyramid Wavefront sensors (PyWFS) \cite{ragazzoni_pupil_1996} have been implemented on current\cite{Pinna_SOUL_2023, Morzinski_2014, Currie_Guyon_2013, jovanovic2019} and planned\cite{hippler2019, davies2016MICADO, thatte2021harmoni, Brandl2021metis, Pinna_ANDES_2024, kasper2021pcs, fitzgerald2022_PSI, crane2018NFARIOS, males2024gmagaox} adaptive optics systems (AO). Even as the emergent standard of the field, the design of the PyWFS has nonlinearities and sensitivity reductions when operating on sky. One of these, Optical Gain (OG), affects the strength of reconstruction between calibration conditions and operations with residuals within the linear regime of the PyWFS. As an effect that scales with seeing strength, OG particularly affects the visible light adaptive optic systems, and will have huge impacts on the corrections possible with larger apertures. Accurately accounting for and compensating for optical gain terms in control and post processing are a priority for current systems in preparation for the PyWFS planned on the next generation 30m class telescopes\cite{esposito_non_2015, deo_assessing_2018, chambouleyron_pyramid_2020}. 

Without optical gain compensation, a deformable mirror (DM) command is reconstructed as a lower value (typically). At first order, this makes an AO system inefficient, resulting in differential residuals in closed loop greater than expected for optimal implementation. Since the WF is underestimated, this limits the accuracy of an open loop reconstruction. This limits the impact of data driven predictive control techniques\cite{haffert_data-driven_2021} in realtime, and point spread function (PSF) reconstruction techniques in post-processing\cite{BeltramoMartin2020}. For many systems, this also hinders non common path corrections if offsets are sent to the same DM as seen by the PyWFS. Essentially, the OG effect hinders other dependent high contrast operations and science goals. 

\begin{figure}
    \centering
    \includegraphics[width=0.80\linewidth]{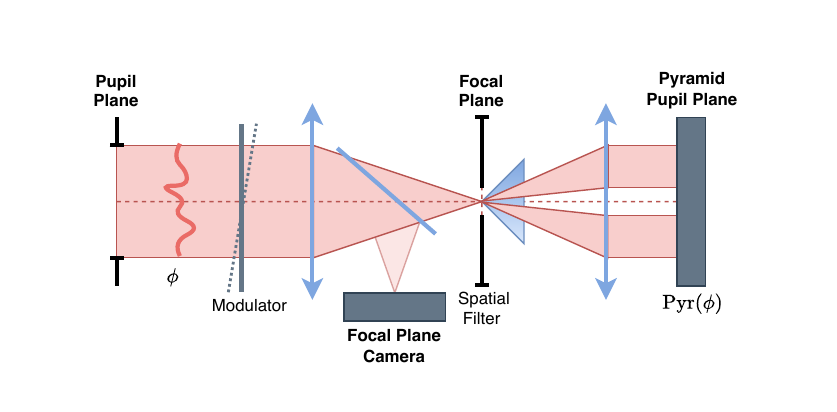}
    \caption{A simplified diagram of the PyWFS architecture. After light is split by upstream beam splitters, collimated light encounters the planar modulator, which imparts a circular tip tilt to the light around the pyramid tip. Light is then focused down to the pyramid tip, split by a pyramid pattern, before being collimated into pupils for the pyramid pupil plane. There is an optional pellicle that sends a fraction of light to a focal plane camera. }
    \label{fig:pywfs_diagram}
\end{figure}

MagAO-X\cite{males2022magaox}, the Magellan Extreme AO system, is a high contrast coronagraphic visible light instrument on the Magellan 6.5m Clay telescope at Las Campanas Observatory (LCO). The system consists of two common path DMs, PyWFS operated at 810nm and at speeds up to 3.6kHz, with a low order WFS run off of rejected coronagraphic light and a non-common path DM. A simplified schematic of the MagAO-X WFS arm can be seen in Figure \ref{fig:pywfs_diagram}. 

In this work we present three different approaches to measuring OG: a full modal diagnostic (section \ref{sec:selfRM}), a focal plane measurement (section \ref{sec:CamTip}), and a simultaneous probe on the pupil plane (section \ref{sec:sparkles}). For each method we describe the methods for data acquisition, the reduction technique, and the datasets collected. In section \ref{sec:discussion} we compare the relative performance of each method and conclude in section \ref{sec:conclusion} with future steps and approaches for MagAO-X. 

\section{Benchmark measurement}
\label{sec:selfRM}

\begin{figure}
    \centering
    \includegraphics[width=0.95\linewidth]{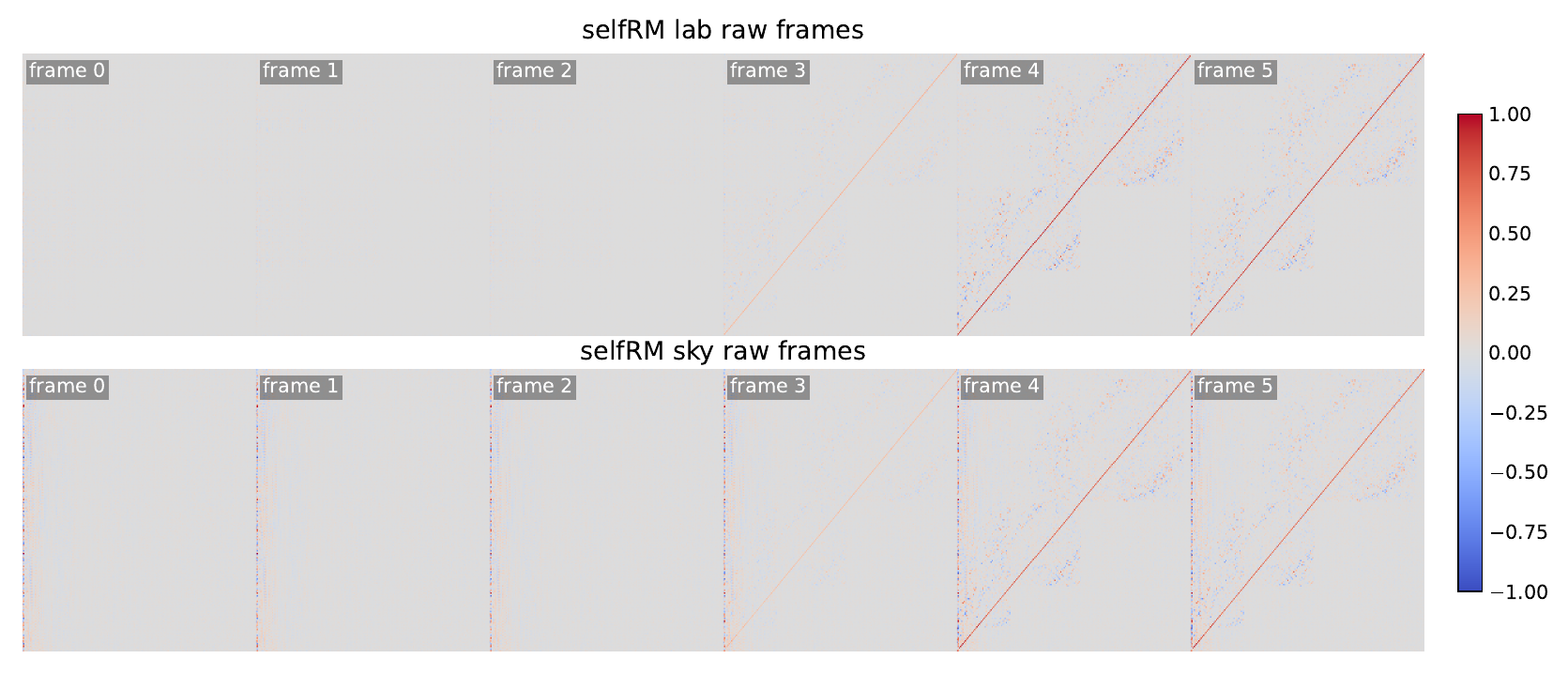}
    \caption{An example of a selfRM matrix in lab conditions (top) and during an observation (bottom). The response is shown for the first 200 modes, for 6 loop iterations. Since the probe is introduced in frame 0 and we have a 2.5 frame delay at 2kHz, we expect the full response to peak 3 full frames after the probe, captured in frame 4.}
    \label{fig:selfRM}
\end{figure}

In previous work\cite{mcewen_-sky_2024} we have described the self response matrix (selfRM) as a mode by mode measurement of optical gain. This differs from a calibration matrix, as we track the full application and degradation of the DM probe as seen by the wavefront sensor. An example of a selfRM for the first 200 modes for the first 5 timesteps of loop iterations (frames) since the probe was applied (frame 0) is shown in Figure \ref{fig:selfRM}. Between a lab calibration and a sky frame the loss in sensitivity is apparent easily on the diagonal modes. 

This technique is mentioned as it is a true measurement of OG, where the modal response is directly measured and the fraction response with wavefront residuals compared to calibration conditions can be calculated. Practically, this is a tool for benchmarking other measures of OG, as it requires probing the DM, adding aberrations and negatively impacting downstream science. For a more detailed account of selfRMs as used on MagAO-X, please refer to previous work\cite{mcewen_-sky_2024}.    

In these previous studies, we discovered the selfRM returned an OG that was unexpectedly flat with respect to mode compared to other studies. In a physical examination of the Pyramid, we found that the spatial filter that came with the pyramid had a diameter of 1.9". When simulating the system with the High Contrast Imaging for Python (HCIPy)\cite{por2018hcipy} package, we found that compared to a diameter of 3.0", it is the 1.9" spatial filter that affects the response of the higher order modes, suppressing the selfRM OG curve to approximately flat past the first 200 modes. 

These selfRM studies allowed us to make two assumptions for our approach going forward on MagAO-X, first, that the optical gain we calculate from modes 200 onwards is relatively constant and can be approximated with a single scalar. Second, that scalar approximation of optical gain has a functional relationship with the performance of the AO system, as it is impacted by WFS residuals. We are interested in finding OG methods that measure this scalar or an equatable value from a functional relation. 

\section{Focal Plane}
\label{sec:CamTip}

MagAO-X is equipped with a pellicle in the WFS path that can be flipped into the beam to send 10\% of the light to a focal plane camera. This gives us an option to observe the pyramid tip PSF independently from measurements on WFS itself. 

    \subsubsection{CamTip: Pyramid Tip camera}
    MagAO-X has a Basler acA640-750$\mu m$ camera positioned past the pellicle. It is a CMOS camera with a 4.8 $\mu m$  square pixel, full resolution of 672 x 512 pixels. At default readout it can readout 751fps, and up to 1724 in a cropped readout. The readnoise has been measured to 8.7 e/s with a comparatively negligible dark current. 
    In the MagAO-X f69 beam, we sample 11 pixels per lambda/D. The measurements described here were taken in a 128x128 FOV, which samples 5.8 $\lambda/D$ in radius, where the modulator for the Pyramid WFS is run at 3 $\lambda/D$. 
    This camera and configuration were not optimized for this project's measurements. Our beam is oversampled and the FOV undersamples fitting error. 

    \subsection{Method for measuring Strehl ratio}

    There have been impressive returns with using a focal plane camera for gain sensing \cite{striffling_optical_2025}. However, the background noise and FOV cut for our camera mean there is not enough fitting error to use our system in this way. Instead, we elect to do a Strehl ratio (SR) measurement, which can be traced to system performance and has a functional relationship to optical gain. 
    We calculate Strehl ratio through a peak to volume measurement, calibrated against a lab measurement. Each frame is dark subtracted, centered via an autocorrelation with a centered lab frame, and then the counts in the inner and outer annuli are calculated. See the diagram below in Figure \ref{fig:CamTip_ex}. The ratio of peak to volume measured during an onsky observation is divided by the same ratio from lab measurements. In this way we normalize out most systematics, but are not able to fully capture the fitting error. 

    \begin{figure}
        \centering
        \includegraphics[width=0.95\linewidth]{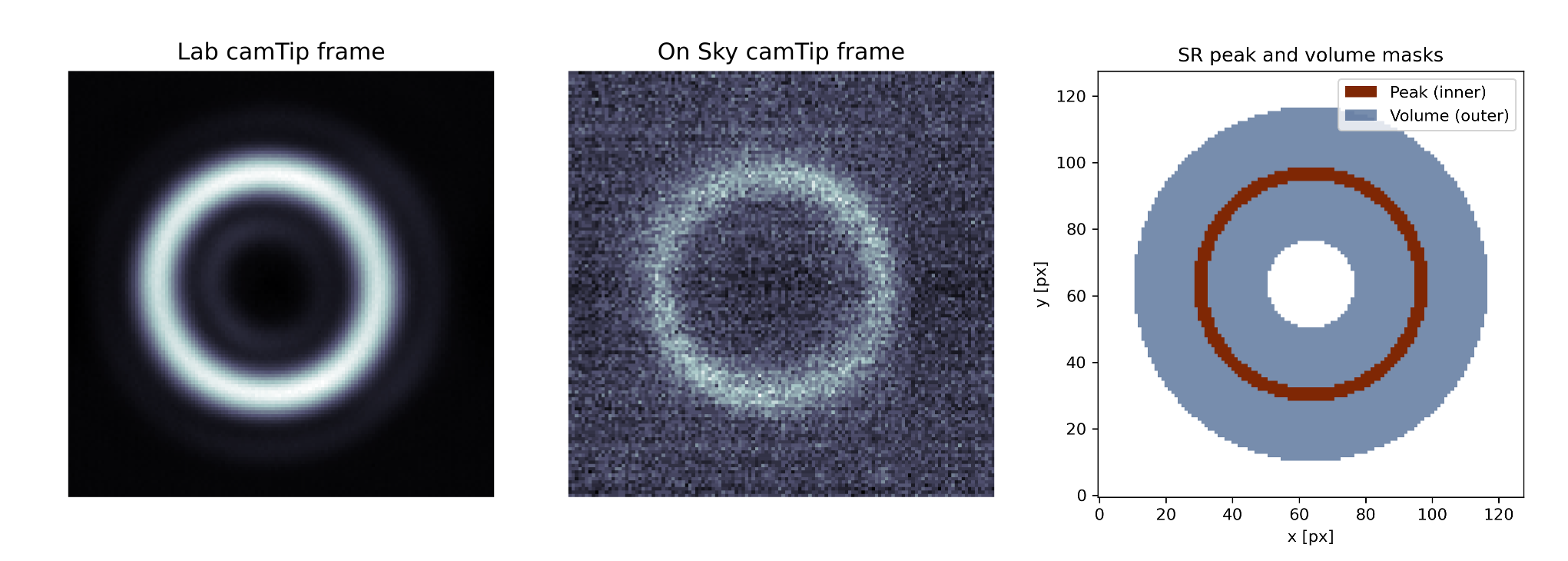}
        \caption{Example frames from the focal plane camera. (Left) Frame from lab source, minimum residual wavefront errors. (Middle) Frame from stellar source, operating through atmospheric residuals. (Right) diagram of showing masks used for defining peak and volume measurements in SR calculation.}
        \label{fig:CamTip_ex}
    \end{figure}

    \subsection{Error budget Strehl ratio}
    We estimate the expected SR at the pyramid tip with an error budget (EB) for comparison to expected system performance\cite{males_ground-based_2018}, detailed in AOSystem source code\cite{males_jaredmalesaosystem_2025}. We use the seeing reported by the site's Differential Image Motion Monitor (DIMM) and the LCO average wind speed (18.7 m/s)\cite{prieto_giant_2010}. We use this EB not for a definitive measure of SR at the pyramid tip, but to investigate any biases and limitations of the focal plane SR estimations. 

    \subsection{SR as an estimator}
    For each of our available datasets that had a corresponding CamTip dataset, we calculated both the SR on the camera frame and estimated the SR with the above error budget. See Table \ref{tab:CamTip_data} in the appendix for a full summary of datasets taken. We have a range of seeing conditions (0.35 - 1.02as at 500nm) and a range of magnitudes (-2.45 to 6.67). While the brightest targets were able to run CamTip at maximum framerates (900-1700/s), the faintest targets ended up needing slower speeds to build up enough signal to noise (10-20 fps typically). One data set series, taken in especially poor seeing (0.86–1.02as) on a middling-brightness target (3.29 magnitude in I band), ran at 1 Hz.  
    
\begin{figure}
        \centering
        \includegraphics[width=0.95\linewidth]{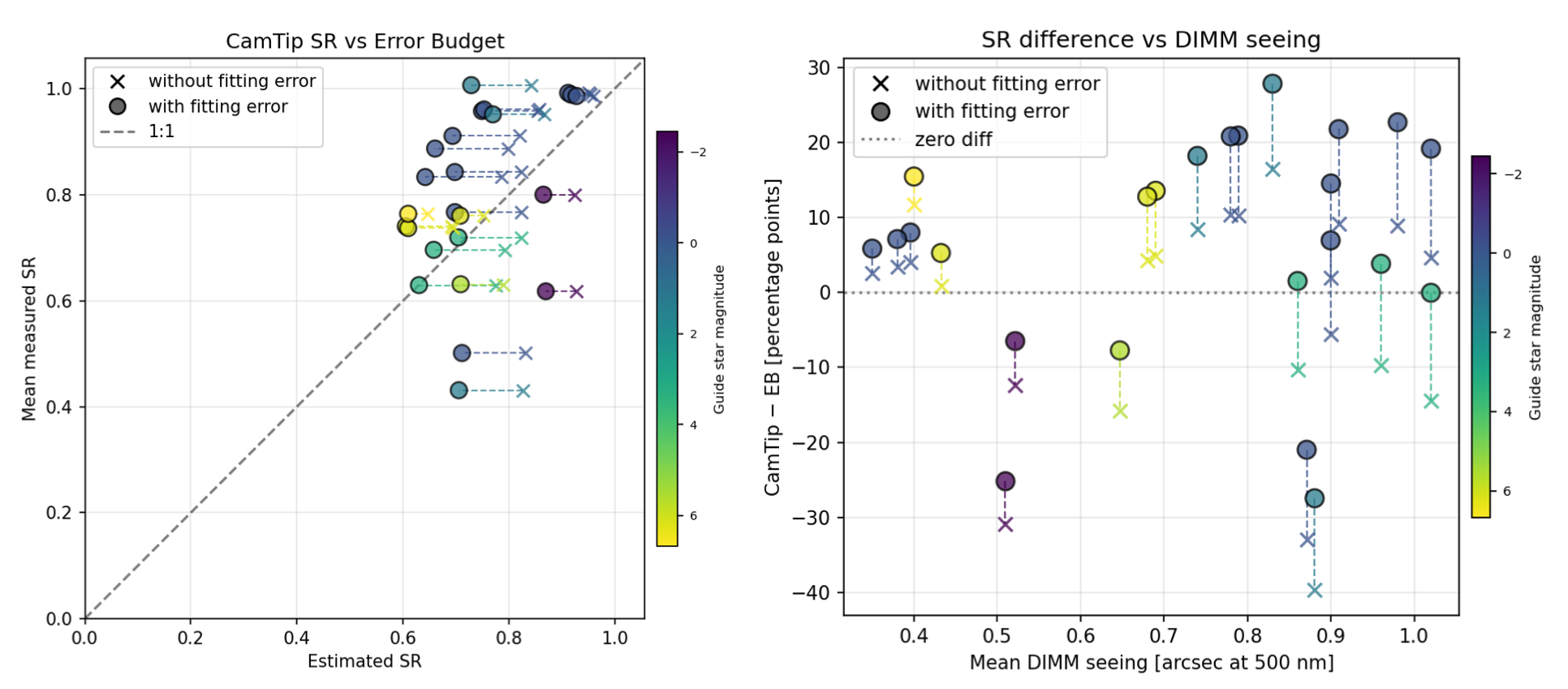}
        \caption{(Left) The estimated SR plotted against the SR measured from the focal plane camera, averaged over the dataset taken. Plotted again with an EB SR that does not contain fitting error terms. (Right) The difference between estimated and predicted error budget plotted against DIMM seeing The recalculated EB SR without fitting error brings about 2/3s of the points closer to the CamTip value, supporting that CamTip cannot accurately capture fitting error.}
        \label{fig:CamTip_summary}
    \end{figure}

    In general, CamTip seems to overestimate compared to the error budget in most datasets as seen in Figure \ref{fig:CamTip_summary}. Given our suspicions that the setup of the focal plane camera cannot accurately measure fitting error, we ran the error budget twice, once with all terms and a second time with fitting error excluded. We display the error budget comparison to the measured SR in Figure \ref{fig:CamTip_summary}, both as a 1:1 plot and their difference compared to the seeing where the measurement was taken. The bright-star outliers, especially at median seeing (0.5-0.7as seeing) are most likely due to saturation on the focal plane. A few targets are well matched to the SR estimate, but considering the slow integration times, this is likely due to a compounding error from added jitter and insufficient fitting error than a single CamTip dataset being accurate. Excluding these outliers, almost every over-estimate from CamTip is somewhat explained with inadequate measurement of fitting error. 

\section{Pupil Plane}
\label{sec:sparkles}

    Incoherent speckles are a pattern of Fourier probes applied to the DM to create PSF copies in the focal plane\cite{jovanovic_artificial_2015}. These are in regular use on MagAO-X for coronagraphic observations, as they provide coronagraphic centering and photometric reference. As these probes are regularly applied, they are available for optical gain measurements without operational changes. The implementation for MagAO-X of this technique closely follows the work done on SCExAO\cite{sahoo_precision_2020}. 
    The incoherent speckles are kept from the loop correction by averaging to zero every two loop iterations with a positive followed by a negative signal at loop speed. They are kept incoherent with the focal-plane speckles by applying a $\pi$ phase shift every two frames. Optical gain tracking with these probes follows work done at LBT\cite{esposito_non_2015} which applied slowly varying probes to estimate OG and rescale the PyWFS reconstructor by its inverse, in order to compensate for non-common-path aberrations (NCPAs).

    \begin{figure}
        \centering
        \includegraphics[width=0.95\linewidth]{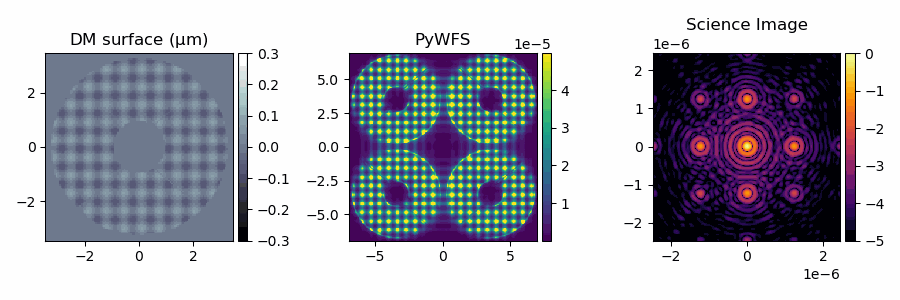}
        \caption{Simulation of applied incoherent speckles. (Left) the probe command as sent to the DM. (Middle) Pattern as seen on the PyWFS. (Right) Science focal plane images showing the PSF replicas from the incoherent speckles.}
        \label{fig:placeholder}
    \end{figure}

\subsection{Calculating Optical Gain from WFS frames}
\label{sec:spark_og_method}

    Since these speckle probes are applied one per loop iteration, the WFS has one frame to sense a pattern out of four. The strength of this pattern can be improved by stacking every fourth frame, repeated for each of the four patterns. However, the WFS integration and DM command are not aligned. In other words, the pattern applied by the DM does not match the frame as seen by the WFS.

    This affected the initial attempts for signal extraction through a simple projection of the speckle patterns onto built up frames.  Initial attempts using these probes simply took the WFS signal as found in lab and projected onto every fourth frame average. The returned signal strength was degraded because the WFS frames integrated over the DM probe mid-transition. The resulting image was not a linear combination of the Fourier probes and the simple projection of Fourier probes was not sufficient.

\begin{figure}
        \centering
        \includegraphics[width=0.95\linewidth]{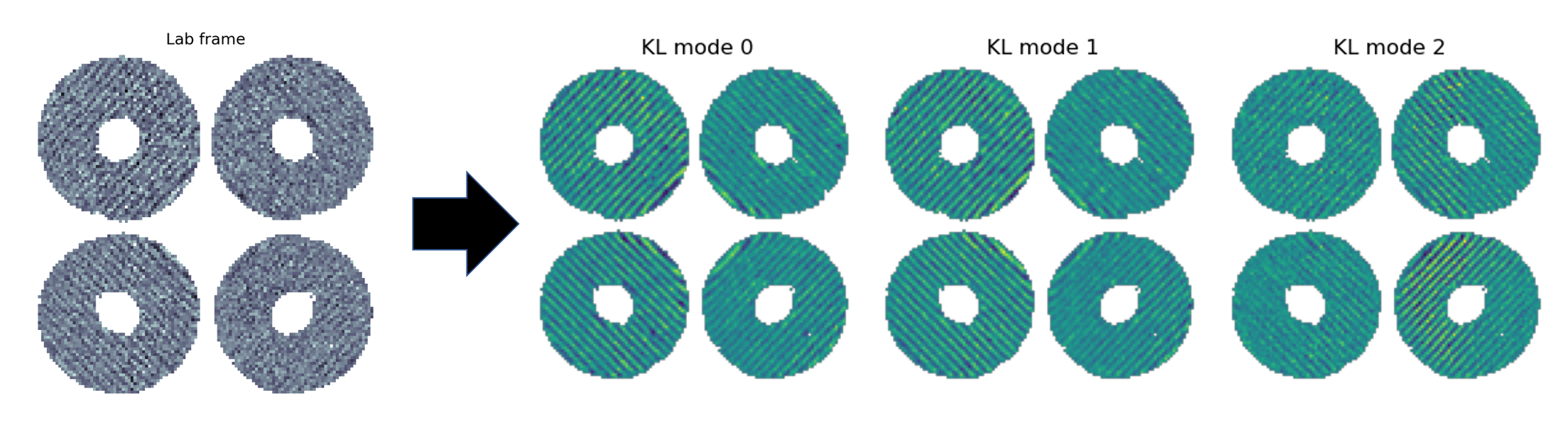}
        \caption{The lab frame Fourier probe image and the three top KL modes from the 2000 frame calibration. The $\pi/2$ phase shift is evident between the first and second mode, with the transition frame in mode 2 showing through variable strength of the probe. }
        \label{fig:KL_images}
    \end{figure}

    Here, we instead collect a dataset in lab conditions and decompose a 2-seconds set using principal component analysis (PCA). From the returned Karhunen–Loève (KL) modes\cite{Karhunen1947, Loeve1978} the top three will always contain the Fourier modes. Typically, the first two are the primary Fourier modes applied, while the third contains the transition information, as seen in Figure \ref{fig:KL_images}.

\subsection{Calibrations}
\label{sec:spark_calibs}

    Each set of likely separations (10-22 $\lambda /D$), strengths (0.01-0.04 $\mu m$) and angle rotations (0-90 degrees in 15 degree increments) was calibrated in lab conditions. These all have been taken with the standard modulation of 3 $\lambda/D$ and at the regular AO loop rate of 2000Hz. While observations have been taken in other configurations, these are standard for good returns on MagAO-X.

\begin{figure}
        \centering
        \includegraphics[width=0.95\linewidth]{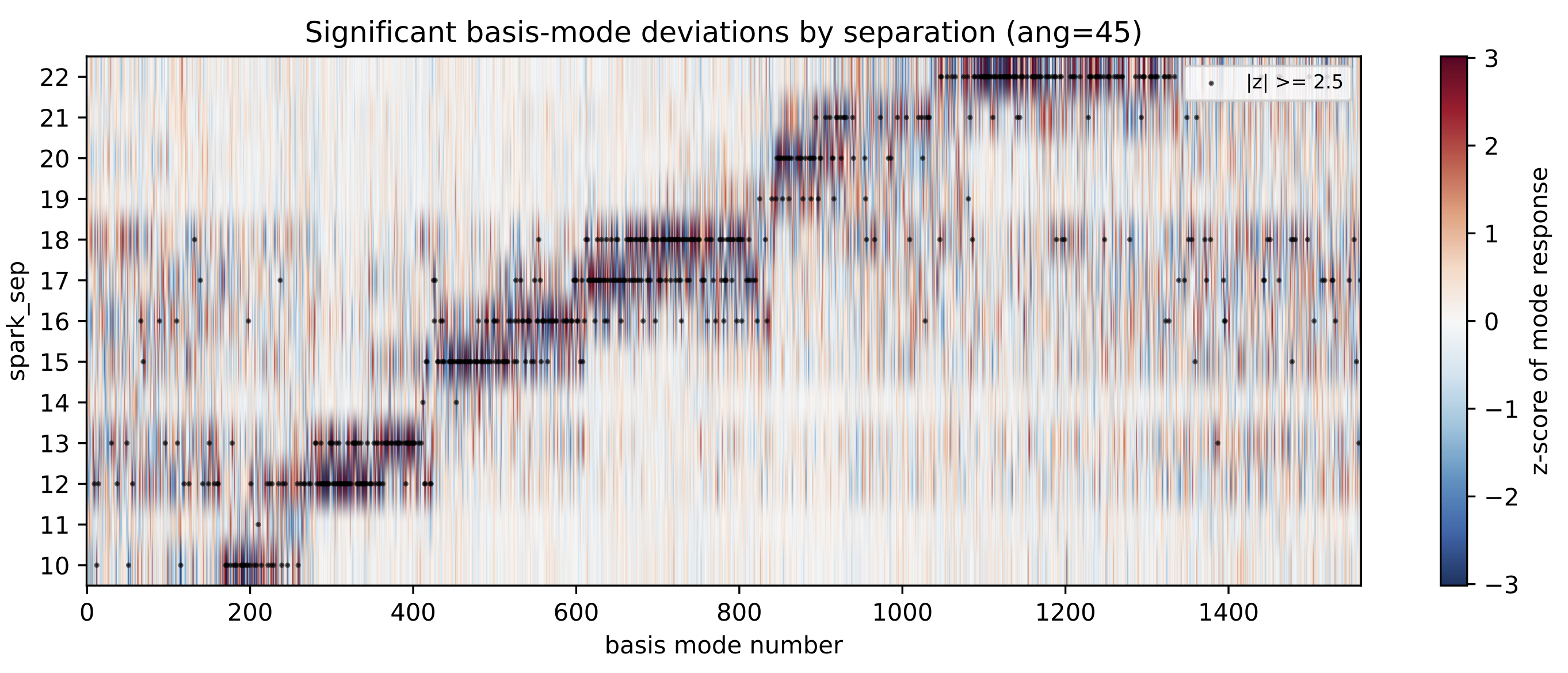}
        \caption{Projecting the probe KL basis values onto the MagAO-X control basis. Z-scores indicate the relative strength of projection relative to the average projection of all speckle basis sets. The basis sets follow the expected behavior with larger separation probing higher order and thus higher frequency modes.}
        \label{fig:spark_projection}
    \end{figure}

    We are interested in how the imposed probes sample the controlling basis of MagAO-X, the basis on which we have selfRM measurements. We are interested in sampling in a range of modes to average over specific spikes, and to sample within the linear regime where the scalar approximation is the most applicable. With lab calibrations of each of the speckle datasets, we project the PCA modes onto the control matrix, determining which modes our system would use to reconstruct the probes. The resulting significance values are shown for a 45 degrees rotated sinusoid in Figure \ref{fig:spark_projection}.

\subsection{Example time series observation}
\label{sec:spark_timeseries}
    
\begin{figure}
        \centering
        \includegraphics[width=0.95\linewidth]{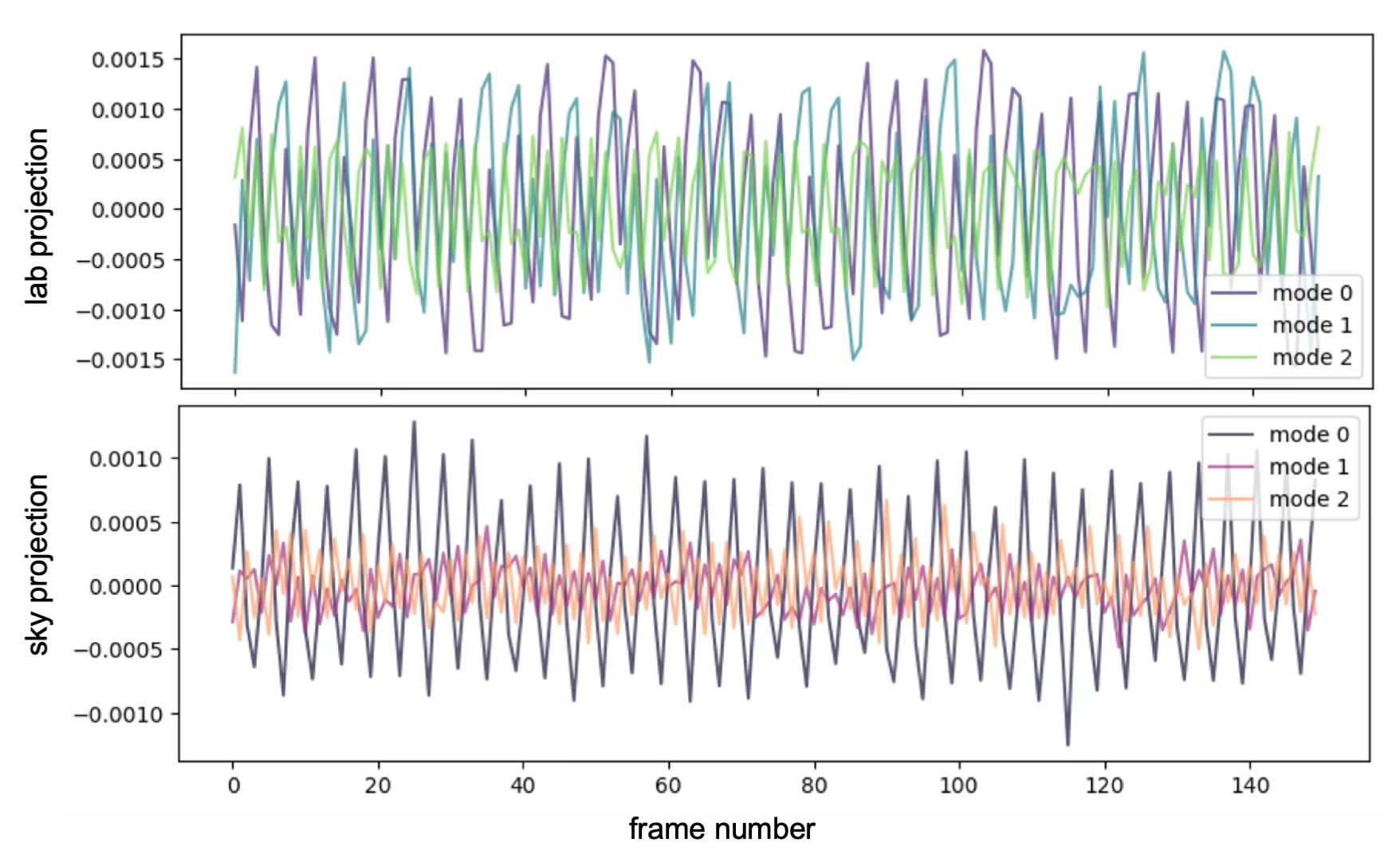}
        \caption{A 150 frame time series of KL mode projections generated from lab frames back onto the same lab frames and onto sky frames. A periodic four frame signal signal is seen clearly for modes 0 and 1 in lab, where mode 2 has a distinct 2 frame pattern, as would be expected for a transition mode. Though the same modes used on sky, these patterns aren't seen as clearly on the sky dataset, which is still being investigated.}
        \label{fig:KL_timeseries}
    \end{figure}

    By projecting each of these KL modes onto individual frames, we trace the cycle of imposed modes, as seen in Figure \ref{fig:KL_timeseries} where the KL modes are projected onto laboratory frames. Though the specific order of the modes can change, we typically see at least two of the three modes on 4 frame cycles, with the third mode on a 2 frame cycle. Applying this same projection on sky frames, we see the expected frequencies but the returns at different amplitudes, indicating that some signals are less visible through turbulence. To average over variations due to seeing and account for the positive negative nature of the probe, we take the RMS of the projection in lab and on sky and divide to get the scaled fraction of the mode on sky. We take the same length RMS as taken in lab, and roll this window over the time series to get the fractional value over the current frame. 

\begin{figure}
        \centering
        \includegraphics[width=0.95\linewidth]{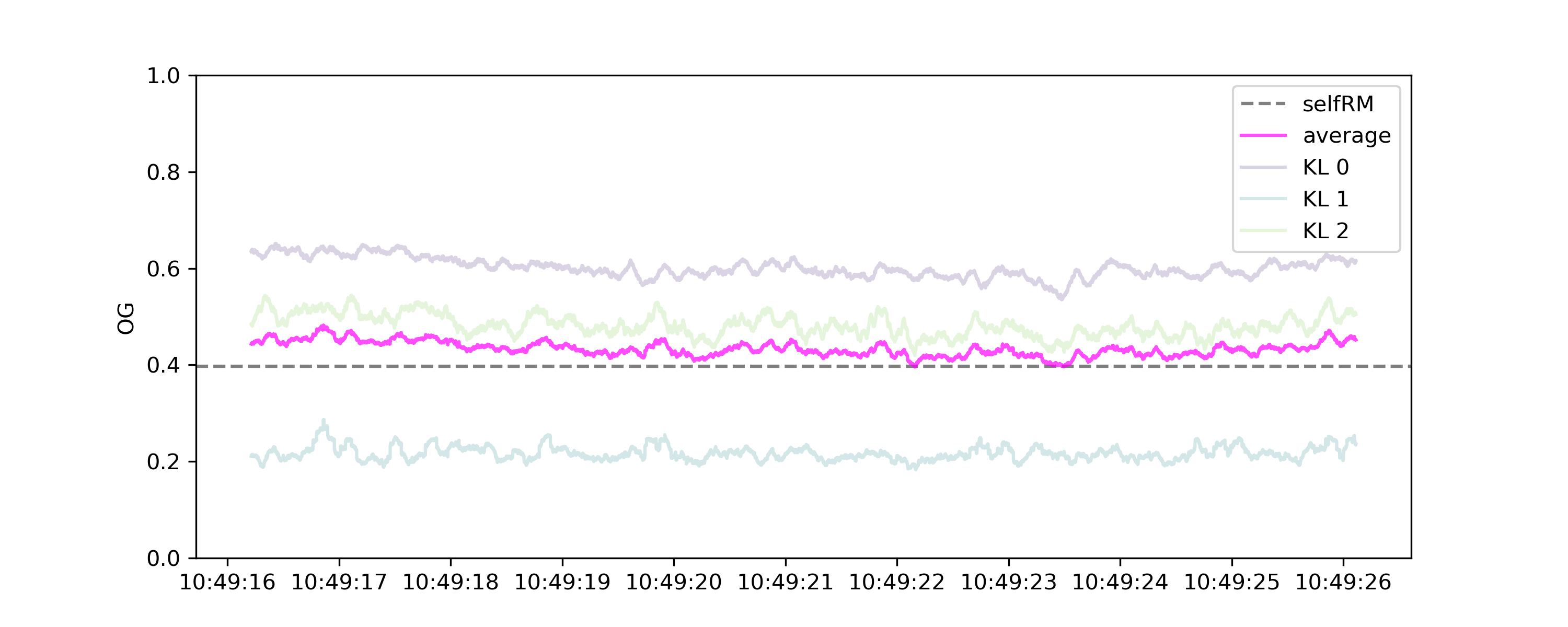}
        \caption{A 10 second timeseries showing the RMS projection for each KL mode, divided by the lab RMS. Plotted also is the average of these values as well as the selfRM average OG as taken just before this dataset was taken. None of the KL modes returns an OG that match the selfRM value well, as expected. The average OG across the KL modes is used as the measurement of optical gain.}
        \label{fig:spark_timeseries}
    \end{figure}

    In Figure \ref{fig:spark_timeseries} the rolling lab-normalized RMS is taken over the three Fourier KL modes. No individual mode matches closely with the OG reported by the selfRM, which is consistent with the failure of a simple projection of modes to describe OG. We expect the actual probe signal to be defined across the basis of the three KL modes. When the KL mode returns are averaged, we see that the signal has a better match to the selfRM OG. We will use the average of the three projected modes as our OG scalar estimate moving forward. 

\subsection{Summary of incoherent speckles observations}
\label{sec:spark_summary}

\begin{figure}
        \centering
        \includegraphics[width=0.95\linewidth]{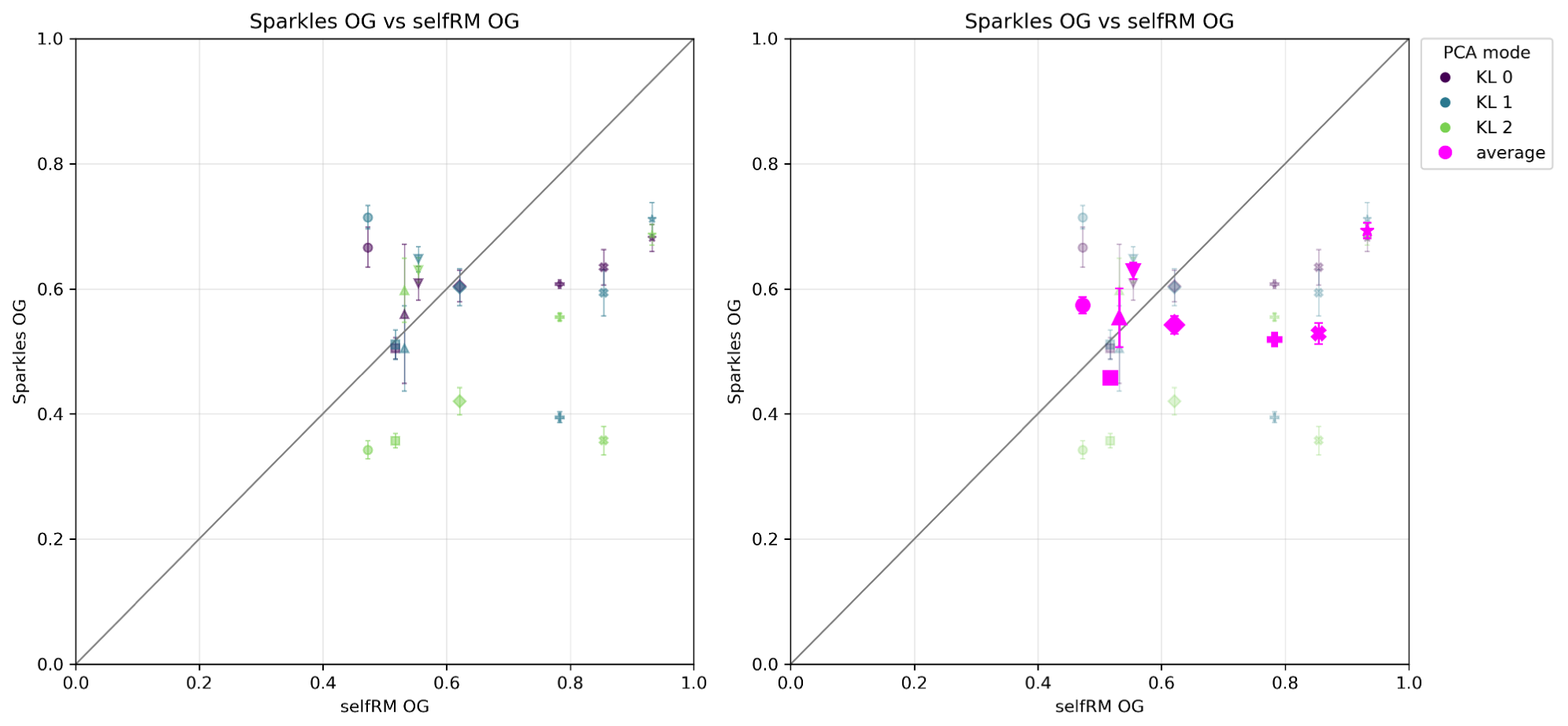}
        \caption{For each selfRM OG measurement each of the KL mode OGs are plotted (left) and for each dataset their average is plotted (right) error bars indicate variance around the average measurement. Though mid range optical gains have reasonable scatter around the unity line with the selfRM measurement, higher optical gains seem to saturate.}
        \label{fig:sparkle_summary}
    \end{figure}

    In the fall of 2025 we took 8 datasets with incoherent speckles and a corresponding selfRM datapoint. For each dataset, we compared the OG measurement in the 5 minutes closest to the selfRM point against the selfRM OG itself, plotted in Figure \ref{fig:sparkle_summary}. The resulting values are the averages of the timeseries shown in Figure \ref{fig:spark_timeseries}. For five of these datasets, the correspondence to selfRM OG is close. For the higher OG values, we see an underestimation of OG. Understanding this underestimation is part of ongoing work.

\section{Method Comparison and discussion}
\label{sec:discussion}

In a series of observations in 2024 and 2025 MagAO-X took a combination of these three measurement tools. These tests were done in parallel to observers' time, and so do not represent a full range of targets and observing conditions, but what targets and conditions were available for this engineering task. SelfRMs provide a way to compare optical gains measured by the loop to estimates of OG from other methods and estimates of system-performance from error budgets. 

The focal plane measurements require inserting a pellicle into the beam, both taking light away from wavefront sensors and potentially adding small additional aberrations from initial calibration. The limited detector sensitivity, coupled with the background noise, led to a low signal-to-noise ratio for targets fainter than 6th magnitude, or under conditions of significant turbulence. In an effort to run faster than the jitter, which does not affect the pyramid WFS measurements, we cropped to a tight FOV that undersamples fitting error. This effect is theoretically calibrated, but in practice saturation and SNR also hinder measurement.

With regard to operational complexity, incoherent speckles are by far the easiest method for passively measuring optical gain. On MagAO-X, this method's OG quantity is always available as long as the applied speckles match one of the lab-calibrated configurations. These values, while still suffering from saturation in some cases, are never outside of the expected range of optical gains.

The selfRMs are still used as a system diagnostic, but come at the cost of time to science, typically around 2 minutes before slewing to another target. CamTip impacts the WFS path and has significant hardware hurdles. Though not fully characterized, the incoherent speckles method is a lower profile modification to the system with most of the reduction process already implemented. 

\section{Conclusions}
\label{sec:conclusion}

In this work we explored three different ways of measuring optical gains or values related to optical gain. Though we have a direct measurement of optical gain in selfRMs, we look towards other methods for low impact ways of measuring realtime optical gain. We found that the focal plane, in which we measure SR to tie to system performance, had significant drawbacks.  In the pupil plane, the incoherent speckles that are already applied are a pre-existing probe that we've calibrated in lab. In practice on the datasets we have available, there seems to be some limitation in OG recovery in the case of high optical gain. 

Future work on this project will aim towards disentangling the limitations in the incoherent speckle method. Once we understand why OG is underestimated, it can be quickly integrated into the control schema of MagAO-X, as it is already being automatically calculated from routines. In follow up tests we can apply not just the singular scalar, but a per modal gain to see if the specificity improves control. 

\acknowledgments 
 
This paper includes data gathered with the 6.5 meter Magellan Telescopes located at Las Campanas
Observatory, Chile. Thank you to the MagAO-X observers and collaborators who gave time for selfRM data collection from their observing time.
We are very grateful for the support from the NSF MRI Award \#1625441 for MagAO-X. The Phase II upgrade program is made possible by the generous support of the Heising-Simons Foundation. The development of CACAO is supported by the National Science Foundation under NSF Award \#2410616. Eden McEwen is supported through the National Science Foundation Graduate Research Fellowship Grant DGE-2137419. 
This research made use of HCIPy, an open-source object-oriented framework written in Python for performing end-to-end simulations of high contrast imaging instruments (Por et al. 2018).

\bibliography{report} 
\bibliographystyle{spiebib} 

\section*{APPENDIX}

\begin{table}[h]
\begin{center}
\begin{tabular}{llrrrlrrr}
\toprule
date & target & mag & FPS & length & DIMM & CamTip SR & EB SR & diff  \\
(UT)  &   & (I) &   & (h:m:s) & (as at 500nm) & (\%) & (\%) & (\%) \\
\midrule
    4-16 & HR1996 & 5.57 & 1724.1 & 9:25 & 0.647 $\pm$ 0.019 & 63.14 $\pm$ 0.01 & 70.87 $\pm$ 0.81 & -7.73 $\pm$ 0.01  \\
    4-16 & betelgeuse & -2.45 & 10.0 & 5:54 & 0.522 $\pm$ 0.006 & 80.03 $\pm$ 0.76 & 86.52 $\pm$ 0.25 & -6.49 $\pm$ 0.76  \\
    4-17 & betelgeuse & -2.45 & 1724.1 & 0:16 & 0.510 $\pm$ 0.000 & 61.81 $\pm$ 0.01 & 86.99 $\pm$ 0.00 & -25.18 $\pm$ 0.01  \\
    4-17 & HD92987 & 6.20 & 435.7 & 35:12 & 0.432 $\pm$ 0.002 & 69.93 $\pm$ 1.18 & 70.80 $\pm$ 0.08 & -0.87 $\pm$ 1.18  \\
    4-17 & alfCen & -0.10 & 925.9 & 0:27 & 0.350 $\pm$ 0.010 & 98.38 $\pm$ 0.04 & 92.78 $\pm$ 0.33 & +5.60 $\pm$ 0.04  \\
    4-17 & alfCen & -0.10 & 925.9 & 1:01 & 0.380 $\pm$ 0.000 & 98.38 $\pm$ 0.04 & 91.77 $\pm$ 0.00 & +6.62 $\pm$ 0.04  \\
    4-17 & alfCen & -0.10 & 925.9 & 6:12 & 0.396 $\pm$ 0.006 & 98.75 $\pm$ 0.04 & 91.22 $\pm$ 0.23 & +7.53 $\pm$ 0.04  \\
    4-17 & HD163296 & 6.67 & 20.0 & 8:09:32 & 0.400 $\pm$ 0.000 & 76.41 $\pm$ 1.63 & 60.98 $\pm$ 0.00 & +15.44 $\pm$ 1.63  \\
    4-21 & HD92987 & 6.20 & 10.0 & 0:22 & 0.680 $\pm$ 0.000 & 73.73 $\pm$ 1.62 & 60.97 $\pm$ 0.00 & +12.76 $\pm$ 1.62  \\
    4-21 & HD92987 & 6.20 & 10.0 & 0:09 & 0.690 $\pm$ 0.000 & 74.08 $\pm$ 1.23 & 60.54 $\pm$ 0.00 & +13.54 $\pm$ 1.23  \\
    4-21 & spica & 1.30 & 925.9 & 45:11 & 0.880 $\pm$ 0.007 & 43.10 $\pm$ 0.71 & 70.54 $\pm$ 0.32 & -27.44 $\pm$ 0.71  \\
    4-21 & spica & 1.30 & 925.9 & 0:10 & 0.830 $\pm$ 0.000 & 98.29 $\pm$ 0.07 & 72.87 $\pm$ 0.00 & +25.41 $\pm$ 0.07  \\
    4-21 & spica & 1.30 & 925.9 & 0:31 & 0.740 $\pm$ 0.000 & 94.61 $\pm$ 0.10 & 76.98 $\pm$ 0.00 & +17.64 $\pm$ 0.10  \\
    4-21 & alfCen & -0.10 & 925.9 & 0:16 & 0.780 $\pm$ 0.000 & 95.96 $\pm$ 0.07 & 75.31 $\pm$ 0.00 & +20.65 $\pm$ 0.07  \\
    4-21 & alfCen & -0.10 & 925.9 & 3:48 & 0.789 $\pm$ 0.013 & 95.72 $\pm$ 0.06 & 74.90 $\pm$ 0.59 & +20.82 $\pm$ 0.06  \\
    4-21 & alfCen & -0.10 & 925.9 & 1:33 & 0.871 $\pm$ 0.012 & 50.14 $\pm$ 0.07 & 71.13 $\pm$ 0.56 & -20.99 $\pm$ 0.07  \\
    4-21 & alfCen & -0.10 & 925.9 & 6:40 & 0.900 $\pm$ 0.011 & 76.73 $\pm$ 0.23 & 69.80 $\pm$ 0.51 & +6.93 $\pm$ 0.23  \\
    4-21 & alfCen & -0.10 & 925.9 & 0:12 & 0.900 $\pm$ 0.000 & 84.29 $\pm$ 0.17 & 69.78 $\pm$ 0.00 & +14.51 $\pm$ 0.17  \\
    4-21 & alfCen & -0.10 & 925.9 & 0:18 & 0.980 $\pm$ 0.040 & 88.71 $\pm$ 0.12 & 66.06 $\pm$ 1.86 & +22.66 $\pm$ 0.12  \\
    4-21 & alfCen & -0.10 & 925.9 & 1:03 & 1.020 $\pm$ 0.000 & 83.35 $\pm$ 0.24 & 64.20 $\pm$ 0.00 & +19.15 $\pm$ 0.24  \\
    4-21 & alfCen & -0.10 & 925.9 & 8:27 & 0.909 $\pm$ 0.006 & 91.10 $\pm$ 0.13 & 69.35 $\pm$ 0.29 & +21.74 $\pm$ 0.13  \\
    4-21 & alfAra & 3.29 & 1.0 & 0:12 & 1.020 $\pm$ 0.000 & 62.98 $\pm$ 3.21 & 63.02 $\pm$ 0.00 & -0.03 $\pm$ 3.21  \\
    4-21 & alfAra & 3.29 & 1.0 & 1:45 & 0.860 $\pm$ 0.029 & 71.98 $\pm$ 0.33 & 70.45 $\pm$ 1.32 & +1.53 $\pm$ 0.33  \\
    4-21 & alfAra & 3.29 & 1.0 & 0:43 & 0.960 $\pm$ 0.013 & 69.63 $\pm$ 1.81 & 65.81 $\pm$ 0.59 & +3.82 $\pm$ 1.81  \\
    \bottomrule
    \\
    \end{tabular}
    \end{center}
    \caption{Datasets taken on CamTip during April 2025 observing run. Error indicates variation over dataset. No change in DIMM seeing or only one sample per time slice will result in no variation for the EB SR. }
    \label{tab:CamTip_data}
\end{table}

\end{document}